
\magnification=1200

\pageno=1
\centerline {\bf AN EXACT MEMBRANE QUANTIZATION  }
\centerline {\bf FROM $W_{\infty}$ SYMMETRY    }
\medskip
\centerline {\bf Carlos Castro }
\centerline {\bf I.A.E.C  }
\centerline {\bf 1407 Alegria}
\centerline {\bf Austin , Texas 78757}
\smallskip

\centerline {\bf March,1995 }
\smallskip
\centerline {\bf ABSTRACT}

An exact quantization of the spherical membrane moving in flat target spacetime
backgrounds is performed.
Crucial ingredients are the exact integrabilty of the $3D~SU(\infty)$
continuous Toda equation and the quasi-finite highest
weight irreducible representations of $W_{\infty}$ algebras. Both continuous
and  discrete energy levels are found.
The latter are found for
periodic-like solutions. Membrane wavefunctionals solutions are found in terms
of Bessel's functions and plausible relations
to singleton field theory are outlined.

\smallskip

\smallskip


 \centerline {\bf {I. Introduction }}
{\vbox {\vskip .5 truein}}

Recently, [1] exact instanton solutions to $D=11$ spherical membranes moving in
flat target spacetime backgrounds were
constructed. The starting point was dimensionally-reduced Super Yang-Mills
theories based on the infinite dimensional
$SU(\infty)$ algebra. The latter algebra is isomorphic to the area-preserving
diffeomorphisms of the sphere. In this fashion the
super Toda molecule equation was recovered preserving one supersymmetry out of
the $N=16$ expected. The expected critical target
spacetime dimensions for the ( super) membrane , $D=27 (11)$ , was closely
related to that of non-critical (super ) $W_{\infty}$
string theories. A BRST analysis revealed that the spectrum of the membrane
must have a relationship to the first unitary
minimal model of a $W_N$ algebra adjoined to a critical $W_N$ string in the
$N\rightarrow \infty$ limit [1]. It is the purpose
of this work to push this connection forward.

In {\bf II} we briefly review the contents of [1] and show the crucial role
that the continous Toda equation, a $D=3$
integrable field theory, [2,3] has in the membrane quantization program. In the
final section we quantize the continous Toda
equation and establish the relationship between this quantization program  and
the construction of highest weight
representations of $W_{\infty}$ algebras [10,11]. This enables us to establish
the connection between the membrane
quantization and representations of $W_{\infty}$ algebras via the continuous
Toda theory. Brief comments about black holes,
singleton-field theory and universal string theory are presented at the
conclusion.

\centerline {\bf II}

Based on the observation that the spherical membrane moving in $D$ spacetime
dimensions, in the light-cone gauge, is essentially
equivalent to a $D-1$ Yang-Mills theory, dimensionally reduced to one time
dimension, of the $SU(\infty)$ group (see Duff [20]
for a review ) we look for solutions of the $D=10$ Yang-Mills equations
(dimensionally-reduced to one temporal dimension).

In [1] we obtained solutions to :

$$\partial_a F_{ab} +[A_a,F_{ab}] =0.~A^\alpha_a T_\alpha \rightarrow A_a( x^b;
q,p).~[ A_a,A_b] \rightarrow \{ A_a,A_b \}_{q,p}.
\eqno (1)$$

with $a,b,...   =8$ being the transverse indices to the membrane ; we performed
the $10=2+8$ split of the original $D=10$ YM
equations. After the dimensional reduction to one dimension  we found that the
following $D=10$ YM potentials, ${\cal A}$ are
one class of solutions to the original equations  :

$${\cal A}_1 =p_1 A_1.~{\cal A}_5 =p_2 A_1. ~{\cal A}_3 =p_1 A_3.{\cal A}_7
=p_2 A_3. \eqno (2a)$$
$${\cal A}_2 =p_1 A_2.~{\cal A}_6 =p_2 A_2.~{\cal A}_0 ={\cal A}_4 ={\cal A}_8
={\cal A}_9 =0.  \eqno (2b)$$

where $p_1,p_2$ are constants and  $A_1,A_2,A_3$ are functions of $x_0,q,p$
only and obey the $SU(\infty)$ Nahm's equations :

$$\epsilon_{ijk}{\partial A_k \over \partial x_0} +\{A_i,A_j \}_{q,p} =0.~i,j,k
=1,2,3. \eqno (3)$$

Nahm's equations were obtained from reductions of Self Dual Yang-Mills
equations in $D=4$. The temporal variable $x_o
=p_1X_0+p_2X_4$. We refer to  [1] for details.

Expanding $A_y =\sum A_{yl}Y_{l,+1}.~A_{\bar y} =\sum A_{{\bar y}l}Y_{l,-1}$;
and $A_3$ in terms of $Y_{l0}$,
the ansatz which allows to recast the $SU(\infty)$ Nahm's equations as a Toda
molecule equation is [1] :

$$\{A_y,A_{\bar y} \} =-i\sum^{\infty}_{l=1}~exp[K_{ll'}\theta_{l'}]Y_{l0}
(\sigma_1,\sigma_2).~A_3
=-\sum^{\infty}_{l=1}{\partial \theta_l\over \partial \tau}.Y_{l0}. \eqno (4)$$

with $A_y ={A_1+iA_2\over \sqrt 2}.~A_{\bar y} ={A_1-iA_2\over \sqrt 2}.$

Hence, Nahm's equations become :

$$-{\partial^2 \theta_l\over \partial \tau^2 } =e^{K_{ll'} \theta_{l'}}.~~
l,l'=1,2,3....\eqno (5)$$

This is the $SU(N)$ Toda molecule equation in Minkowski form. The $\theta_l$
are the Toda fields where $SU(2)$ has been embedded
minimally into $SU(N)$. $K_{ll'}$ is the Cartan matrix which in the continuum
limit becomes :$\delta''(t-t')$ [2,3]. The
continuum limit of (5) is

$$-{\partial^2 \theta (\tau,t) \over \partial \tau^2 }
=exp~[\int~dt'\delta''(t-t') \theta (\tau,t')]. \eqno (6)$$

Or in alternative form :

$$-{\partial^2 \Psi(\tau,t) \over \partial \tau^2 } = \int~\delta''
(t-t')exp[\Psi (\tau,t')]~dt' ={\partial^2 e^{\Psi }\over
\partial t^2}. \eqno (7)$$

if one sets $K_{ll'}\theta_{l'} =\Psi_l$. The last two equations are the
dimensional reduction of the $3D\rightarrow 2D$
continuous Toda equation given by Leznov and Saveliev :

$${\partial^2 u\over \partial \tau^2} =-{\partial^2 e^u\over \partial
t^2}.~i\tau =r=z+{\bar z}. \eqno (8)$$

This equation was obtained by Boyer,Finley and Plebanski, as rotational Killing
symmetry reductions of Self-Dual Gravity in
$D=4$.

In [1] we established the correspondence between the target space-times of
non-critical $W_{\infty}$ strings and that of
membranes in $D=27$ dimensions. The supersymmetric case was also discussed and
$D=11$ was retrieved. It was shown in [12] that
the effective induced action of $W_N$ gravity in the conformal gauge takes the
form of a Toda action for the scalar fields and
the $W_N$ currents take the familiar free field form. The same action can be
obtained from a constrained $WZNW$ model. Each of
these Toda actions posseses a $W_N$ symmetry. The authors [12,13] coupled $W_N$
matter to $W_N$ gravity in the conformal gauge,
and integrating out the matter fields, they arrived at the induced effective
action which was precisely the same as the Toda
action. Non-critical $W_N$ strings are constructed the same way. The matter and
Liouville sector of the $W_N$ algebra can be
realized in terms of $N-1$ scalars, $\phi_k,\sigma_k$ repectively. These
realizations in general have background charges which
are fixed by the Miura transformations [5,6]. The non-critical string is
characterized by the central charges of the matter and
Liouville sectors, $c_m,c_L$. To achieve a nilpotent BRST operator these
central charges must satisfy :

$$c_m+c_L =-c_{gh} =2\sum^N_{s=2} (6s^2-6s+1) =2(N-1)(2N^2+2N+1). \eqno (9)$$

In the $N\rightarrow \infty$ limit a zeta function regularization yields
$c_m+c_L =-2$.

We were able to show that the critical membrane background in $D=27$ was the
same as that of a non-critical $W_{\infty}$ string
background if one adjoined the first unitary minimal model of the $W_N$ algebra
to that of a critical $W_N$ string spectrum in
the $N\rightarrow \infty$ limit. In particular :

$$c_{eff} =D=1-12x^2 =(1-12x^2_o) +c_{m_o} = 26 -(1-{6\over (N+1)(N+2)})
+2{N-1\over N+2} \Rightarrow D=27. \eqno (10)$$
The value for the total central charge of the matter sector is $c_m =2+{1\over
24}$ after a zeta function regularization.
That of the Liouville sector is $c_L =-4-{1\over 24}$. These values bear an
important connection to the notion of Unifying $W$
algebras [4]. It happens that when the central charges have for values at
$c(n)={2(n-1)\over n+2};{2(1-2n)\over
n-2};-1-3n$, there exists a Unifying Quantum Casimir $W$ algebra :

$${\cal W}{\cal A}_{n-1} \leftrightarrow {\cal W} (2,3,4,5) \sim {sl {\hat
(2,R)}_n \over {\hat U(1)}}. \eqno (11)$$

in the sense that these algebras truncate at degenerate values of the central
charge to a smaller algebra.

We see that the value $c_m$ after regularisation corresponds to the central
charge of the first unitary minimal model of
$WA_{n-1}$ after $n$ is analytically continued to a negative value $n=-146
\Rightarrow 2(n-1)/(n+2) =2+1/24.$
The value of $c_L$ does not correspond to a minimal model but it also
corresponds to a special value of $c$ where the
$WA_{n-1}$ algebra truncates to that of the unifying coset : $c(n)=
2(1-2n)/(n-2) =-4 -1/24$ for $n=146$.
In virtue of the
quantum equivalence equivalence between negative rank $A_{n-1}$ Lie algebras
with $n=146\rightarrow -n=-146; c=2+{1\over 24}$
and  ${\cal W} (2,3,4,5)\sim W_{\infty}$ at $c=2$; i.e. the Hilbert spaces are
isomorphic ,
we can study the spectrum of non-critical ${\cal W} (2,3,4,5)$ strings and
claim that it ought to give very relevant
information concerning the membrane's spectrum. Unfortunately, non-critical
${\cal W} (2,3,4,5)$ strings are prohibitely
complicated. One just needs to look into the cohomology of ordinary critical
$W_{2,s}$ strings to realize this [5].
Nevertheless there is a way in which one can circumvent this problem. The
answer lies in the integrabilty property of the
continuous Toda equation [2,3] and the recently constructed quasi-finite
highest weight irreducible representations of
$W_{1+\infty}, W_{\infty}$ algebras [10,11]. For this purpose one needs to
compute explicitly the value of the coupling constant
appearing in the exponential potential of the Quantum Toda theory [1]. The
latter is conformally invariant and the conformally
improved stress energy tensor obeys a Virasoro algebra with an adjustable
central charge whose value depends on the coupling
constant $\beta$ appearing in the potential [15]. This value coincides
precisely with the one obtained from a Quantum
Drinfeld-Sokolov reduction of the $SL(N,R)$ Kac-Moody algebra at level $k$ :

$$\beta ={1\over \sqrt {k+N}}.~~~c(\beta) =(N-1) -12 |\beta \rho -(1/\beta)
{\tilde \rho})|^2. \eqno (12)$$

where $\rho,{\tilde \rho}$ are the Weyl vectors of the (dual) $A_N$ algebra. We
see [1,12] that one can now relate the value of
the background charge $x$ in (10)  and $\beta$ when
$k=-\infty.~N=\infty.~k+N=constant$ :

$$2x^2=(-13/3) =({1\over \sqrt {k+N}} -\sqrt {k+N})^2 = (\beta -{1\over \beta}
)^2 \Rightarrow \beta^2 = {-7+(-)\sqrt 13 \over
6}.    .\eqno (13)$$

so $\beta$ is purely imaginary. This should not concern us.
There exist integrable field theories known as Affine Toda theories whose
coupling is imaginary but posseses soliton solutions
with real energy and momentum [9].

\centerline {\bf { III }}

{\vbox {\vskip .5 truein}}

Having reviewed the essential results of [1] permits us to look for classical
solutions to the
continous Toda equation and to implement the Quantization program presented in
[2].
The general solution to ( 8 ) depending on two
variables , say $r\equiv z_+ +z_-$ and $t$ (not to be confused with time ) was
given by [2].
The solution is determined by two
arbitrary functions , $\varphi (t)$ and $d(t)$. It is :

$$exp[-x(r,t)]=exp[-x_o(r,t)]\{1+\sum_{n> 1}~(-1)^n\sum_\omega~\int
\int....exp[r\sum_{m=1}^n~\varphi (t_m)] \Pi~dt_m d(t_m)$$
$$[\sum_{p=m}^n~\varphi
(t_p)]^{-1}[\sum_{q=m}^n~\varphi (t_{\omega (q)})]^{-1}. [\epsilon_m
(\omega)\delta
(t-t_m)-\sum_{l=1}^{m-1}~\delta''(t_l-t_m)\theta
[\omega^{-1}(m)-\omega^{-1}(l)]]\}. \eqno (14)$$
with :
$\rho_o=\partial^2x_o/\partial t^2 =r\varphi (t) +ln~d(t).$ This defines the
boundary values of the solution
$x(r,t)$ in the asymptotic region $r\rightarrow \infty $.

$\theta$ is the Heaviside step-function.
$\omega$ is any permutation of the indices from $[2..........n] \rightarrow
[j_2,..........j_n]$.

$\omega (1)\equiv 1.$ $\epsilon_m (\omega)$ is a numerical coefficient. See [2]
for details.

An expansion of (14) yields :
$$exp[-x]=exp[-x_o]\{1-\mu +{1\over 2}\mu^2+........\}.\eqno (15)$$
where :
$$\mu \equiv {d(t)exp[r\varphi (t)]\over \varphi^2}. \eqno  (16)$$

The solution to the Quantum $A_{\infty}$  ( continous) Toda chain can be
obtained by taking the continuum limit of
the general solution to the finite nonperiodic Toda chain associated with the
Lie algebra
$A_N$ in the $N\rightarrow \infty$ limit. This is performed by taking the
continuum limit of eqs-(82-86) of [3] :

$$\varphi_i \rightarrow x_o(r,t).~\psi_{j_s} \rightarrow \partial^2
x_o/\partial t^2_s =r\varphi (t_s) +ln~d(t_s). \eqno (17)$$
In the $r\rightarrow \infty$ limit the latter tends to $r\varphi (t_s)$.

$$\sum_{j_1j_2...j_n} \rightarrow \int~\int~....dt_1 dt_2.......dt_n.~{\cal
P}^1 \rightarrow [\sum \varphi (t_p) +O(h)]^{-1}.
{\cal P}^2 \rightarrow [\sum \varphi (t_{\omega (q)}) +O(h)]^{-1}.\eqno (18)$$

Therefore, one just has to write down the quantum corrections
to the  two factors $[\sum~\varphi]^{-1}$ of eq-(14).
One must replace the first factor by a summation from $p=m$ to $p=n$ of terms
like :
$$[\varphi (t_p) +O(h)] \rightarrow \varphi (t_p)-(i h/2) [{1\over w
(t_p)}]_{,t_pt_p} -i h\sum^n_{l=p+1}
[{1\over w (t_l)}]_{,t_lt_l}. \eqno (19)$$

and the second factor  by  a summation from $q=m$ to $q=n$ of terms like :

$$[\varphi (t_{\omega (q)}) +O(h)] \rightarrow [eq~(19) : ~p\rightarrow \omega
(q)] +
 i h
-i h \sum_{l=1}^{q-1}[{1\over w(t_{\omega (l)})}]_{,t_{\omega (l)} t_{\omega
(l)}}$$
$$ +i h
\sum_{l=q+1}^n[{1\over w(t_{\omega (l)})}]_{,t_{\omega (l)} t_{\omega (l)}}
\eqno (20)$$

where $w (t)$ is a positive function that is the continuum limit of eqs-(30,34)
of [3]. What one has done is to replace :

$${\hat k}_{j_mj_l}\equiv {k_{j_mj_l}\over w_{j_m}} \rightarrow \int~dt_m
{\delta'' (t_m-t_l)\over w (t_m)} =
[{1\over w (t_l)}]_{,t_lt_l}. ...\eqno (21)$$
in all the equations in the continuum limit. One has smeared out the delta
functions in the denominators of (14) using the
function $w(t)$. If one had not smeared out the delta functions one would have
encountered ill-defined expressions.
{}From now on we set $ h=1$.

These are the quantum corrections to the classical solution $\rho = \partial^2
x/\partial t^2 $ where $x(r,t )$ is given
in (14). These  are the continuum limits of eqs-(82-86) of [3].  It is
important to realize that one must not add quantum
corrections to the $\varphi, d(t) $ appearing in the terms $exp[r\sum \varphi]$
and $x_o$ of (14). The former are two arbitrary
functions which parametrize the space of solutions. It is $\rho$ and $x$ which
acquire quantum corrections given by (19,20)
through the $w(t)$ terms and, as such, $\rho, x$ should be seen as quantum
operators acting on the Hilbert space of states.
Upon quantization, $h$ appears and associated with Planck's constant a new
parametric function has to appear : $w(t)$.     One
has to incorporate also the coupling constant $\beta$ in all of the equations .
This is achieved by rescaling the continuous
Cartan matrix by a factor of $\beta$ so that  $\partial^2 x/\partial t^2$ and
$\partial^2 x_o/\partial t^2$ are rescaled  by a
factor of $\beta$; i.e. $r\varphi$ acquires a factor of $\beta$ and
$d(t)\rightarrow d(t)^\beta$.
 Since $\beta$ is pure imaginary, for convergence purposes in the
$r=\infty$ region we must have that $\beta \varphi <0\Rightarrow \varphi
=i\varphi$ also. In the rest of this section we will
work without the $\beta$ factors and only reinsert them at the end of the
calculations. There is nothing unphysical about this
value of $\beta$ as we said earlier.

One of the integrals of motion is the energy. The continuos Toda chain is an
exact integrable system in the sense
that it posesses an infinite number of functionally independent integrals of
motion : $I_n (p,\rho)$ in involution. i.e.
The Poisson brackets  amongst $I_n,I_m$ is zero. Since these are integrals of
motion, they do not depend on $r$. These
integrals can be evaluated most easily in the asymptotic region $r\rightarrow
\infty$. This was performed in [2] for the case
that  $\varphi (t)$ was a negative real valued function which simplified the
calculations.
For this reason the energy
eigenvalue given in [2] must now be  rescaled by a factor of $\beta^2$  :

$$E=\beta^2\int^{2\pi}_0~dt (\int^t~dt'\varphi (t'))^2. \eqno (22
)$$

where we have chosen the range of the $t$ integration to be $[0,2\pi]$. Since
$\beta\varphi <0 \Rightarrow \beta^2\varphi^2 >0$
and the energy is positive. We insist, once more, that $t$ is a parameter which
is not the physical time and that $\varphi (t)$
does not acquire quantum corrections. The latter integral (22) is the
eigenvalue of the Hamiltonian which is one of the Casimir
operators for the irreducible representations of $A_N$ in the $N\rightarrow
\infty$ limit.

We can borrow now the results by [10,11] on the quasi-finite highest weight
irreducible representations of $W_{1+\infty} $
and $W_{\infty}$ algebras. The latter is a subalgebra of the former.  For each
highest weight state,$|\lambda >$ parametrized by
a complex number $\lambda$ the authors [10,11] constructed  representations
consisting of a finite number of states at each
energy level by succesive application of ladder-like operators. A suitable
differential constraint on the generating function
$\Delta (x)$ for the highest weights $\Delta^\lambda_k$ of the representations
was necessary in order to ensure that, indeed,
one has a finite number of states at each level. The highest weight states are
defined :

$$W(z^n D^k) |\lambda> =0.~n\ge 1.k\ge 0.~~W(D^k) |\lambda>=\Delta^\lambda_k
|\lambda > .~k\ge 0. \eqno (23)$$

The $W_{1+\infty}$ algebras can be defined as central extensions of the Lie
algebra of differential operators on the circle.
$D\equiv zd/dz.~n\epsilon {\cal Z}$ and $k$ is a positive integer. The
generators of the $W_{1+\infty}$ algebra are denoted by
$W(z^n D^k)$ and the $W_{\infty} $ generators  are obtained from the former :
${\tilde W} (z^n D^k) =W (z^n D^{k+1})$; where
$(n,k \epsilon {\cal Z}.~k\ge 0)$. (There is no spin one current).
The generating function $\Delta (x)$ for the weights is :

$$\Delta (x) = \sum_{k=0}^{k=\infty} \Delta^\lambda_k~{x^k\over k!}. \eqno
(24)$$
where we denoted explicitly the $\lambda$ dependence as a reminder that we are
referring to the highest weight state $|\lambda
>$ and satisfies the differential equation required for quasi-finiteness :

$$b(d/dx)[(e^x-1)\Delta (x) +C] =0.~ b(w)
=\Pi~(w-\lambda_i)^{m_i}.~\lambda_i\not= \lambda_j. \eqno (25)$$

$b(w)$ is the characteristic polynomial. $C$ is the central charge and the
solution is :

$$\Delta (x) ={\sum_{i=1}^K~p_i(x)e^{\lambda_i x} -C\over e^x-1}.   \eqno
(26)$$
The generating function for the $W_{\infty}$ case is ${\tilde \Delta} (x)
=(d/dx) \Delta (x)$ and the central charge is
$c=-2C$.

The Verma module is spanned by the states :

$$|v_\lambda>
=W(z^{-n_1}D^{k_1})W(z^{-n_2}D^{k_2})...........W(z^{-n_m}D^{k_m})|\lambda >.
\eqno (27)$$

The energy level is $\sum_{i=1}^{i=m}~n_i$. For further details we refer to
[10,11]. Highest weight unitary representations for
the $W_{\infty}$ algebra obtained from field realizations with central charge
$c=2$ were constructed in [10].

The weights associated with the highest weight state $|\lambda >$ will be
obtained from the expansion in (24).
In particular, the "energy" operator acting on $|\lambda >$ will be  :

$$W(D)|\lambda > =\Delta^\lambda_1 |\lambda >. \eqno (28)$$
$L_o = -W(D)$ counts the energy level :$[L_o, W(z^n D^k)] =-nW(z^n D^k)$.

As an example we can use for $\Delta (x)$ the one obtained in the free-field
realization by free fermions or {\bf bc} ghosts
[10]

$$ \Delta (x) =C{e^{\lambda x}-1\over e^x -1} \Rightarrow \partial
\Delta/\partial \lambda =C{xe^{\lambda x}\over e^x -1}.
\eqno (29)$$
The second term is the generating function for the Bernoulli polynomials :

$${xe^{\lambda x}\over e^x -1} = 1+(\lambda -1/2)x +(\lambda^2-\lambda
+1/6){x^2\over 2!} + (\lambda^3 -3/2
\lambda^2+1/2\lambda){x^3\over 3!} +.........\eqno (30)$$

Integrating (30) with respect to $\lambda $ yields back :

$$\Delta (x) =C{e^{\lambda x}-1\over e^x -1} =\sum_{k=0}~\Delta_k{x^k\over k!}.
\eqno (31)$$

The first few weights (modulo a factor of $C$) are then :

$$\Delta_0= \lambda.~\Delta_1 = (1/2) (\lambda^2 -\lambda).~\Delta_2 =
(1/3)\lambda^3 -(1/2)\lambda^2 +(1/6) \lambda.....\eqno
(32)$$

The generating function for the $W_{\infty}$ case is ${\tilde \Delta} (x)
={d\Delta (x)\over dx}\Rightarrow {\tilde
\Delta}^\lambda_k =\Delta^\lambda_{k+1}$.

${\tilde \Delta}^\lambda_1=\Delta^\lambda_2$ is the weight associated with the
"energy" operator. In the ordinary string, $W_2$
algebra, the Hamiltonian is related to the Virasoro generator, $H=L_o$ and
states are built in by applying the ladder-like
operators to the highest weight state, the "vacuum". In the
$W_{1+\infty},W_{\infty}$ case it is not longer true that the
Hamiltonian  (52) can be written exactly in terms of the zero modes of the
$W_2$ generator, once the realization of the
$W_{\infty}$ algebra is given  in terms of the continous Toda field given by
[2] :

$$ {\tilde W}^+_2  =\int^{t_o}~dt_1~\int^{t_1}~dt_2~exp[-\Theta (z,{\bar
z};t_1)]{\partial\over \partial
z}exp[\Theta (z,{\bar z};t_1)-\Theta (z,{\bar z};t_2)] {\partial\over \partial
z} exp [\Theta (z,{\bar z};t_2)]. \eqno (33)$$

The chiral generator has the form $W^+_h [\partial \rho/\partial
z....\partial^h \rho /\partial z^h]$ [2]
and the similar expression for the antichiral generator $ W^-_h$ is obtained by
replacing $\partial_z\rightarrow
\partial_{{\bar z}}$ in (33).
After a dimensional reduction from $D=3\rightarrow D=2$ is taken, $r=z+{\bar
z}$, one has :

$${\tilde W}_2 (r,t_o) =\int^{t_o}~dt_1~\int^{t_1}~dt_2~exp[-\rho
(r,t_1)]{\partial\over \partial r}exp[\rho (r,t_1)-\rho
(r,t_2)] {\partial\over \partial r} exp [\rho (r,t_2)]. \eqno (34)$$

When $\rho
(r,t)$ is quantized in eq-(19,20) it becomes an operator, ${\hat \rho}(r,t)$,
acting on a suitable Hilbert space of states, say
$|\rho>$, and in order to evaluate (34) one needs to perform the highly
complicated Operator Product Expansion between the
operators ${\hat \rho} (r,t_1),{\hat \rho} (r,t_2)$. Since these are no longer
free fields it is no longer trivial to compute
per example the operator products :

$${\partial \rho \over \partial r}.e^\rho.~~e^{\rho
(r,t_1)}.e^{\rho(r,t_2)}.....\eqno (35)$$

Quantization deforms the classical
$w_{\infty}$ algebra into $W_{\infty}$ [17,18]. For a proof that the
$W_{\infty}$ algebra is the Moyal bracket deformation of
the $w_{\infty}$ see [18]. Later  in [19] we were able to construct the
non-linear ${\hat W}_{\infty}$ algebras from
non-linear integrable deformations of Self Dual Gravity in $D=4$.
Since the $w_{\infty}$ algebra has been effectively quantized the
expectation value of the ${\tilde W}_2$ operator at tree level,  $<\rho
|{\tilde W}_2 |\rho>$, is related to the ${\tilde W}_2
(classical)$  given by (34).
One can evaluate all expressions in the $r=\infty$ limit ( and set $d(t) =1$
for convenience.
The expectation value $<\rho|{\hat W}_2 ({\hat \rho}) |\rho>(\varphi (t))$
gives in the $r=\infty$ limit, after the dimensional reduction and after using
the asymptotic limits :

$${\partial \rho \over \partial r} =\varphi.~~~{\partial^2 \rho \over \partial
r^2} = {\partial^2 e^\rho \over \partial
t^2}\rightarrow 0.~r\rightarrow \infty\eqno (36)$$

$$lim_{r\rightarrow \infty }<\rho|{\hat W}_2|\rho> = \int^{t_o}dt_1 \varphi
(t_1) \int^{t_o}dt_1 \varphi (t_1).
\eqno (37)$$

after the normalization condition is chosen :

$$<\rho'|\rho>=\delta (\rho'-\rho).~<\rho|\rho> =1               \eqno (38) $$.
We notice that eq-(37) is  the same as the integrand (22); so integrating (37)
with respect to $t_o$ yields the energy.
It is useful to recall the results from ordinary $2D$ conformal field theory :
given the  holomorphic current generator of
two-dimensional conformal transformations, $T (z)=W_2(z)$, the mode expansion
is :

$$W_2 (z) =\sum_m~W^m_2 z^{-m-2}\Rightarrow W^m_2 =\oint~{dz\over 2\pi i}
z^{m+2-1}W_2 (z). \eqno (39)$$
the closed integration contour encloses the origin. When the closed contour
surrounds
$z=\infty$. This requires performing the conformal map $z\rightarrow (1/z)$ and
replacing :

$$z\rightarrow (1/z).~dz\rightarrow (-dz/z^2).~W_2(z) \rightarrow (-1/z^2)^2
W_2 (1/z).  \eqno (40)$$
in the integrand.

There is also a $1-1$ correspondence between local fields and states in the
Hilbert space :

$$|\phi> \leftrightarrow lim_{z,{\bar z}\rightarrow 0} {\hat \phi} (z,{\bar z})
|0>. \eqno (41)$$

This is usually referred as the $|in>$ state. A conformal transformation
$z\rightarrow 1/z; {\bar z} \rightarrow 1/{\bar z}$:
defines the $<out|$ state at $z=\infty$

$$<out| =lim_{z,{\bar z}\rightarrow 0} [(-1/z^2)^h (-1/{\bar z}^2)^{\bar h}
{\hat \phi} (1/z,1/{\bar z}) |0>]^+. \eqno (42)$$

where $h,{\bar h}$ are the conformal weights of the field $\phi (z,{\bar z})$.

The analog of eqs-(42) is to consider the state parametrized by $\varphi (t)$ :
$$|\rho>_{\varphi (t)} =lim_{ r\rightarrow \infty}~| \rho (r,t)>\equiv |\rho
(out)>.~
   |\rho>_{-\varphi (t)} =lim_{ r\rightarrow -\infty}~| \rho (r,t)>\equiv |\rho
(in)>.               \eqno (43)$$

since the continuous Toda equation is symmetric under $r\rightarrow
-r\Rightarrow \rho (-r,t)$ is also a solution and it's
obtained from (14) by setting $\varphi \rightarrow -\varphi$ to ensure
convergence at $r\rightarrow -\infty$.

The state $|\rho>$ is parametrized in terms of $\varphi$ and for this reason
one should always write it as
$|\rho>_\varphi$ . What is required now is to establish the correspondence (a
functor) between the representation space realized
in terms of the continuous Toda field and that representation ( the Verma
module) built from the highest weight $|\lambda>$

$$<\lambda| {\tilde W} (D) |\lambda>={\tilde \Delta}^\lambda_1 \equiv
\Delta^\lambda_2  \leftrightarrow
<\rho|{\hat W}_2[{\hat \rho}(r,t)|\rho>.\eqno (44)$$

What is required then is to integrate with
respect to $t_o$, to extract the zero mode piece of the ${\tilde W}_2$ operator
via a contour integral around the origin ,
  :

$$ {\tilde \Delta}^\lambda_1  = <W^0_2 > \leftrightarrow \int^{2\pi}_0~dt_o~
<\rho|\oint{ dz\over 2\pi i }\oint{d{\bar z}\over
2\pi i}  z{\bar z} {\hat W}_2 [\rho(z,{\bar z},t)] |\rho>. \eqno (45)$$

the contour integral also could be performed around infinity : $z=0 \rightarrow
1/z
=\infty$ if one wishes.  The expectation value of  : $<\rho|{\hat W}_2 |\rho>$
in the dimensionally-reduced
case, depends on  $r=z+{\bar z}\Rightarrow  <\rho|{\hat W}_2|\rho> (r,t)$. The
latter is a function which
can be expanded in powers of $z+{\bar z} :\sum a_n (z+{\bar z})^{-n}$ and the
integral (45) can be computed.

Rigorously speaking one must also
include the realization of the anti-chiral algebra ${\bar W}_{1+\infty}$
[10,11] in terms of ${\bar W} ({\bar z}^n {\bar
D}^k)$ which yields the weights ${\bar \Delta}^{\bar \lambda}_k$. Thus the
r.h.s of (45) involves both types of weights.

Also required is to introduce a family
of functions $\varphi_\lambda (t)$ parametrized by $\lambda$; i.e to each
$|\lambda> \leftrightarrow |\varphi_\lambda>$ so
that  $<\rho|{\hat W}|\rho> (r,t)$ is functionally dependent on
$\varphi_\lambda (t)$. Eq-(45) is an integral equation
relating  ${\tilde \Delta}^\lambda_1$ to the family of functions
$\varphi_\lambda (t)$ linking in this way the realizations in
terms of the continuous Toda field and the highest weight $W_{\infty}$
irrepresentations.

The l.h.s of (45) corresponds to the  action of $ {\tilde W}(D)$ on $
|\lambda>$ in the language of [10,11] . The r.h.s of (45
) corresponds indeed to extract the
zero mode part of the  ${\tilde W}_2$ generator, in the realization of the
$W_{\infty}$ algebra in terms of the continuous Toda
field, after the dimensional reduction $z+{\bar z} =r$ has taken place. The
other weights, $ {\tilde \Delta}^\lambda_k$ and the
anti-chiral ones, are also obtained from the zero modes of Saveliev's
realization of the chiral and antichiral $W_{\infty}$
algebras,

$$W^+_h [\partial \rho/\partial z....\partial^h \rho /\partial z^h].~W^-_h
[\partial_z \rho \rightarrow \partial_{{\bar z}}\rho].~\partial W^+/\partial
{\bar z} =0.~\partial W^-/\partial  z =0\eqno
(46) $$.

$$\int^{2\pi}_0dt_o <\rho|\oint~{dz\over 2\pi i}\oint {d{\bar z}\over 2\pi i}
z^{h-1}{\bar z}^{{\bar h}-1}   W_h |\rho>
\leftrightarrow <\lambda|{\tilde W}(D^k) |\lambda > ={\tilde \Delta}^\lambda_k
. \eqno (47)$$

where $k\ge 1.~<\rho| W_h |\rho> (r,t)$ will depend on $(\varphi_\lambda (t))$
after the dimensional reduction takes place.

Thus Eqs-(45,47) are the equations we were looking for.

Therefore, to conclude,  eqs- (45,47) are the eigenvalue equations which
determines the very intricate relationship between
$\varphi_\lambda (t)$
and the weights ${\tilde \Delta}^\lambda_k$ . Given a quasi-finite
highest-weight irreducible representation; i.e. given the set
${\tilde \Delta}^\lambda (x)\Rightarrow \Delta^\lambda_k;C,b(w),\chi....$ one
can from eq-(45,47) determine $\varphi_\lambda (t)$
as a family of functions parametrized by $\lambda$. It is essential to maintain
that $\varphi (t)< 0$ in order to use the
asymptotic $r=\infty$ expression for the energy (22). Since $\lambda$ is a
continuous parameter the energy spectrum (22)  is
continuous in general. Below we will study a simple
case when one has a discrete spectrum characterized by the positive integers
$n\ge 0$.
It is important to realize that (37)  is  equal to the expression for the
classical energy
density (22) not only at  $r=\infty$ but for other values of $r$
so that :

$${\hat H} [W_2 [{\hat \rho}(\varphi_\lambda (t), d_\lambda(t)) ] |\rho > = E
(\varphi_\lambda) |\rho >.
\eqno (48)$$

we recall that $E(\lambda)$ is not always given by eq-(22). The latter equation
 is valid only for  a very special case when
the function $\varphi (t)$ is  real and negative.

The functions $\varphi (t)<0$ in [2] were taken to belong to the space of
trigonometric polynomials in the circle. One may expand
:

$$\varphi_\lambda (t) =\sum_{m=0}~A_m (\lambda) cos~(mt) +B_m (\lambda )
sin~(mt). \eqno (49) $$
and may take $A_m, B_m$ to be functions of the $\lambda$ parameter
characterizing the representations (like the weights). If this is the case,
having an infinite family of functions in
$\lambda,~{\tilde \Delta}^\lambda_k.~k=1,2......$, the integral equations
(45,47) for $h=2,3,4........$,
will be sufficient to specify  $A_m(\lambda),B_m(\lambda).~m=0,1,2......$. In
this fashion one can determine
$\varphi_\lambda (t)$ and establish the $|\lambda>\rightarrow
|\varphi_\lambda>$ correspondence. Therefore,
plugging (49) into (45,47) determines
the coefficients  $A_m,B_m$ as functions of $\lambda$ when $h=2,3,.....$.
Eqs-(45,47) are difficult to solve in general. There is a special case when we
can solve it.

Below we will study a simple
case when one has a discrete spectrum characterized by the positive integers
$n\ge 0$. We shall restore now the coupling
$\beta^2<0$ given in (13) . A simple fact which allows for the possibility of
discrete energy states is to use the
analogy of the Bohr-Sommerfield quantization condition for periodic system. It
occurs  if one opts to choose for the quantity
$exp[\beta \varphi (t) r]\equiv exp[i\Omega r]$ which appears in (14); $\Omega$
is the frecuency parameter ( a constant ). If
the dynamical system is periodic in the variable $r$
with periods $2\pi /\Omega$, a way to quantize the values of $\Omega$ in units
of $n$ is to recur to the Bohr-Sommerfield
quantization condition for a periodic orbit :
$$ J=\oint~pdq =nh       \eqno (50a)$$

which reflects the fact that upon emission of a quanta of energy
$(h/2\pi)\Omega$ the change in the enery level as a function of
$n$ is  [16] (set $h/2\pi =1$) :

$$  \partial E/\partial n =\Omega ={2\over 3} (2\pi)^3 \Omega \partial \Omega
/\partial n.\Rightarrow $$

$$\Omega (n) ={3\over 2 (2\pi)^3 } n. \eqno (50b)$$
Hence the energy is

$$E={3\over 4}(2\pi)^{-3} n^2.    \eqno (51)$$
which is reminiscent of the rotational energy levels $E\sim h^2l(l+1)$ of a
rotor in terms of the angular momentum quantum
numbers $l=0,1,....$. In order to have a proper match of dimensions it is
required to insert the membrane tension as it happens
with the string.

Saveliev [2] chose the $\varphi (t)$ in (14) to be negative real functions to
assure that the
potential term in the Hamiltonian vanished at $r\rightarrow \infty$ and arrived
at (22).
In  case that the functions $\varphi (t)$ are no longer $<0$; i.e when $\beta
\varphi r$ is no longer a real valued quantity
$<0$,  the asymptotic formula (22) will no longer hold and one will be forced
to perform the very complicated integral !

$${\cal H} = \int~dt[-{1\over 2}\beta^2(\partial p/\partial t)^2 +({m^2\over
\beta^2}) exp~[\beta\partial^2x/\partial t^2 ]
]. \eqno (52)$$

where $p= \beta \partial x/\partial r$ is the generalized momentum
corresponding to $\rho \equiv \beta\partial^2x/\partial
t^2,$ and
 $\mu^2 \equiv ({m^2\over \beta^2})$ is the perturbation theory expansion
parameter discussed in [3]. Without loss of
generality it can be set to one. Nevertheless, eqs-(50,51) are still valid. One
just needs to evaluate the
Hamiltonian at $\Omega r=2\pi p$ where $p$ is a very large integer
$p\rightarrow \infty$ and take $d(t)=0$ in (15,16) :

$$exp[\partial^2 x/\partial t^2]\rightarrow d(t)exp[i2\pi p] =0.~(\partial
p/\partial t)^2\rightarrow (\int \varphi
dt')^2...$$   recovering (22) once again.

Are there zero energy solutions ?. If one naively set $\varphi (t) \equiv 0$ in
(14) or set $n=0$ in (51) one would get a zero
energy. However eqs-(14,15) for the most part  will be singular and this would
be unacceptable. One way  zero
energy states could be obtained is by choosing $\varphi (t), d(t)$
appropriately so that (52) is zero. Since one has one
equation and two functions to vary pressumably there should be an infinite
number of solutions of zero energy.

Solving (48) is analogous to solving an Schroedinger-like equation.
Concentrating on the case that $\varphi (t) <0 $;
 the wave functional is defined :
$\Psi [\rho,t]\equiv <\rho|\Psi>$ where the state $|\rho>_{\varphi}$ has an
explicit dependence on $\varphi$ which also depends
on $\lambda$ as shown in (45,47). Upon replacing (in units $ h/2\pi =1$)

$\partial p/\partial t \rightarrow i(\partial/\partial t \delta/\delta \rho)$
as an operator acting on the $\Psi$, the time-independent Schroedinger-like
equation for the wave functional becomes :

$$[(\partial^2/\partial t^2 \delta^2/\delta \rho^2) +exp~\rho ] \Psi [\rho,t] =
{E (\varphi)\over 2\pi} \Psi [\rho,t].
\eqno (53)  $$
One could have for an eigenvalue-density, $\gamma (t)$, in the r.h.s of (53)
other values than the particular constant  $E/2\pi$
if one wishes. In this case the solutions are more complex.
The action functional is :

$$\int dt\int~{\cal D}\rho dr \Psi^+(i{\partial \over \partial r} -{\cal
H})\Psi. \eqno (54)$$

${\cal D}\rho$ is the functional integration measure; $r$ is the variable
linked with the physical time
and the on-shell condition is just (53).
This is the second-quantization of the physical quantities. $\rho (r,t)$ has
already been first-quantized in (19,20).
One should not interpret $\Psi$ as a probabilty amplitude but as a field
operator ( ''membrane'' field) which creates a
continuous Toda field in a given quantum state $|\rho>_{\varphi}$ associated
with the classical configuration configuration given
by eq-(14) in terms of $\varphi_{\lambda} (t)$.

One can expand $|\Psi>$ in an infinite dimensional basis spanned by the Verma
module  (27) associated with the state
$|\lambda>$. Given a vector $v_\lambda \epsilon  {\cal V}$ ( Verma module)  one
has :

$$|\Psi > = \sum_{v_\lambda}~<v_\lambda ||\Psi > |v_\lambda >.~v_\lambda
\epsilon {\cal V}    \eqno (55)$$

This is very reminiscent of the string-field $\Phi [X (\sigma )] =<X||\Phi
(x_o)>$ where $x_o$ is the center of mass coordinate
of the string and the state $|\Phi (x_o)>$ is comprised of an infinite array of
point fields :

$$|\Phi (x_o)> =\phi (x_o) |0> +A_\mu (x_o) a^{\mu +}_1 |0> +g_{\mu\nu}a^{\mu
+}_1 a^{\nu +}_1 |0> +.....\eqno (56)$$

where the first field is the tachyon, the second is the massless Maxwell, the
third is the massive graviton....
The oscillators play the role of ladder-like operators acting on the
''vacuum''$|0>$ in the same manner that the Verma module is
generated from the highest weight state $|\lambda>$ by succesive application of
a string of $W(z^{-n}D^k)$ operators acting on
$|\lambda>$.
The state $|\rho (r,t)>_{\varphi}$
It is the relative of the string state $|X(\sigma_1,\sigma_2)>$ whereas $|\Psi
>$ is the relative of the string field state
$|\Phi >$. The Schroedinger-like equation is of the form :

$$[\partial_t^2 \partial_y^2 +e^y] \Psi (y,t) =E\Psi (y,t). \eqno. (57)$$

A change of variables :$x=2e^{y/2}$ converts (57) into  Bessel's equation after
one sets $\Psi (y,t) =e^t\Phi (y)$ or
equal to $e^{-t}\Phi (y)$  :

$$  (x^2\partial^2_{x^2} +x\partial_x +x^2-4E)\Phi (x)=0. \eqno (58)$$

and whose solution is :$\Phi (x) =c_1{\cal J}_\nu (x) +c_2 {\cal J}_{-\nu} (x)$
where $\nu \equiv 2\sqrt E$ and $c_1,c_2$ constants.

The wavefunctional is then :

$$\Psi [\rho,t] = \int~dr  e^{t}[c_1{\cal J}_\nu (2e^{\rho (r,t)/2}) +c_2{\cal
J}_{-\nu} (2e^{\rho (r,t)/2})]  \eqno (59)$$
or the other solution involving $e^{-t}...$.

One may notice that  discrete energy level (in suitable units such as
$\nu=2\sqrt E=n$) solutions are possible. Earlier
we saw in (51) that $E(n)\sim n^2$ so $\sqrt E \sim n$. Therefore setting
$2\alpha e^{y/2\alpha}=x$ where $\alpha$ is a
suitable constant allows to readjust $\nu =2{\sqrt \alpha E}=n$. The
Bessel functions will have nodes at very specific points . The solutions in
this case will be given in terms of ${\cal J}_n$ and
the modified Bessel function of the second kind, $K_n$. These solutions are
tightly connected with the boundary conditions of
the wave-functional.

It has been argued that a four dimensional
anti-deSitter spacetime, $AdS_4$, whose boundary is  $S^{2}\times S^1$, could
be realized as a membrane at the end of the
universe. In particular, singleton field theory can be described on the
boundary of $AdS_4$ where singletons are the most
fundamental representations of the de Sitter groups [20]. Moreover, on purely
kinematical grounds, infinitely many massless
states of all spins (massless in the anti-de-Sitter sense) can be constructed
out of just two singletons ( preons). In
particular, the $d=4~N=8$ Supersingleton field theory formulated on the
boundary of $AdS_4$ bears a connection with the
supermembrane moving on  $AdS_4\times S^7$ . The rigid $OSp(8|4)$ symmetry acts
as the superconformal group on the boundary
$S^{2}\times S^1$ [20].  In view of this it is important to study if there is
any connection between the wave-functional
behaviour at the boundaries and singleton field theory. The supersymmetric Toda
equation has also been discussed by [2,3].
Roughly speaking, a membrane is comprised of an infinite number of strings.
Thus the membrane can be seen as a coherent state of
an infinite number of strings . This is reminiscent of the Sine-Gordon soliton
being the fundamental fermion of the massive
Thirring model, a quantum lump [16]. The lowest fermion-antifermion bound state
(soliton-antisoliton doublet) is the fundamental
meson of Sine-Gordon theory. Higher level states are built from excitations of
the former in the same way that infinitely many
massless states can be built from just two singletons.

Perhaps the most relevant physical applications of the membrane quantization
program will be in the behaviour of black hole
horizons [8]. These have been described in terms of a dynamical surface whose
quantum dynamics is precisely that of a
relativistic membrane. Thermodynamical properties like the entropy and
temperature of the black hole were derived in agreement
with the standard results. Results for the level structure of black holes were
given. A ''principal'' series of levels was found
corresponding to the quantization of the area of the horizon in units of the
''area quantum'' :$A=nA_o.~A_o =8\pi$. From each
level of this principal series starts a quasi-continuum of levels due to the
membrane's excitations.

Secondly, the ordinary bosonic string has been found to be a special vacua of
the $N=1 $ superstring [14]. It
appears that there is a whole hierarchy of string theories : $w_2$ string is a
particular vacua of the $w_3$ string and so
forth......If this is indeed correct one has then that the (super) membrane,
viewed  as  noncritical $W_{\infty}$ string theory,
is, in this sense, the universal space of string theory.
The fact, advocated by many, that a Higgs symmetry-breakdown-mechanism of the
infinite number
of massless states of the membrane generates the infinite massive string
spectrum fits within this description.

Finally, we hope that the essential role that Self Dual $SU(\infty)$ Yang-Mills
theory has played in the origins of the
membrane-Toda theory, will shed more light into the origin of duality in string
theory [7].

\smallskip

ACKNOWLEDGEMENTS. We thank M.V. Saveliev for many helpful suggestions
concerning the quantization program.
To Bob Murray for refreshing to us
the appropriate change of variables in connection to Bessel's differential
equation and to the Towne family for their kind and
warm hospitality in Austin, Texas.

 \smallskip

\centerline {\bf REFERENCES}

1. C. Castro : "$D=11$ Supermembrane Instantons, $W_{\infty}$ Strings and

the Super Toda Molecule " IAEC-12-94.

submitted to the Jour. of Chaos, Solitons and Fractals. hepth-lanl-9412160.

2-M.V. Saveliev : Theor. Math. Physics {\bf vol. 92}. no.3 (1992) 457.

3.A.N.Leznov, M.V.Saveliev, I.A.Fedoseev : Sov. J. Part. Nucl. {\bf 16} no.1
(1985) 81.

A.N.Leznov, M.V.Saveliev : "Group Theoretical Methods for Integration of
Nonlinear Dynamical Systems " Nauka, Moscow, 1985.

4.R.Blumenhagen, W. Eholzer, A. Honecker, K. Hornfeck, R. Hubel :" Unifying $W$
algebras".

Bonn-TH-94-01 April-94. hepth-lanl-9404113. Phys. Letts. {\bf B 332} ( 1994)
51.

5.H.Lu, C.N. Pope, X.J. Wang: Int. J. Mod. Phys. Lett. {\bf A9} (1994) 1527.

H.Lu, C.N. Pope, X.J. Wang, S.C. Zhao : "Critical and Non-Critical $W_{2,4}$
Strings".

CTP-TAMU-70-93. hepth-lanl-9311084.

H.Lu, C.N. Pope, K. Thielemans, X.J. Wang, S.C. Zhao : "Quantising Higher-Spin
String Theories "

CTP-TAMU-24-94. hepth-lanl-9410005.

6. E. Bergshoeff, H. Boonstra, S. Panda, M. de Roo : Nucl. Phys. {\bf B 411}
(1994)  717.

7. M.Duff, R. Minasan :''Putting String/String Duality to the Test''

CTP-TAMU-16/94.  hepth-lanl-9406198.

8. M. Maggiore :''Black Holes as Quantum Membranes : A Path Integral Approach
''

hepth-lanl-9404172.

9. M.A.C. Kneippe, D.I. Olive : Nucl.Phys. {\bf B 408} (1993) 565.

10. H.Awata, M.Fukuma, Y.Matsuo, S.Odake :"Representation Theory of the
$W_{1+\infty}$ Algebra".

RIMS-990 Kyoto preprint , Aug.1994.

S.Odake : Int.J.Mod.Phys. {\bf A7} no.25 (12) 6339.

11. V.Kac, A. Radul : Comm. Math. Phys. {\bf 157} (1993) 429.

12.J. de Boer : "Extended Conformal Symmetry in Non-Critical String Theory" .

Doctoral Thesis. University of Utrecht, Holland. (1993).

13. J.de Boer, J. Goeree : Nucl. Phys. {\bf B 381} (1992) 329.

14. N. Berkovits, C.Vafa :Mod. Phys. Letters {\bf A9} (1994) 653.

15.P. Bouwknegt, K. Schouetens : Phys. Reports {\bf 223} (1993) 183.

16.S. Coleman : "Aspects of Symmetry " Selected Erice Lectures.

Cambridge Univ. Press. (1989). Chapter 6,page 239.

17.C.Pope, L.Romans, X. Shen : Phys. Lett. {\bf B 236} (1990)173.

18. I.Bakas, B.Khesin, E.Kiritsis : Comm. Math. Phys. {\bf 151} (1993) 233.

D. Fairlie, J. Nuyts : Comm. Math. Phys. {\bf 134} (1990) 413.

19. C.Castro :"Non-linear $W_{\infty}$ Algebras from Non-linear Integrable
Deformations of

 Self Dual Gravity" hepth-lanl-9409197.

20. M.Duff : Class. Quant. Grav. {\bf 5} (1988) 189.

E.Bergshoeff, A.Salam, E.Sezgin ,Y.Tanii : Nucl. Phys. {\bf B 305} (1988) 497.

E.Bergshoeff, E.Sezgin ,P.Townsend  :  Phys. Letts. {\bf B 189} (1987) 75.

\bye